\documentclass[a4paper,12pt]{article}
\usepackage{amssymb}
\usepackage[latin1]{inputenc}     

\newcommand{\beq}{\begin{equation} }
\newcommand{\eqq}{\end{equation} }
\newcommand{\cuad}{{\sqcap\kern-.68em\sqcup}}

\newtheorem{remark}{Remark}[section]
\newcommand{\bremark}{\begin{remark} \em}
\newcommand{\eremark}{\end{remark} }

\def\beeq{\begin{equation}}
\def\eeq{\end{equation}}
\newcommand{\begeqaet}{\begin{eqnarray*}}
\newcommand{\eneqaet}{\end{eqnarray*}}

\let\Section=\section
\def\section{\setcounter{equation}{0}\Section}
\newtheorem{Lem}{Lemma}[section]
\newtheorem{Thm}{Theorem}[section]
\newtheorem{Def}{Definition}[section]
\newtheorem{Prop}{Proposition}[section]
\newtheorem{Remark}{Remark}[section]

\begin{document}
\begin{center}{\bf\Large Existence and symmetric result for Liouville-Weyl fractional nonlinear Schr\"odinger equation }\medskip

\bigskip

\bigskip

{C\'esar Torres}

 Departamento Acad\'emico de Matem\'aticas\\ 
 Universidad Nacional de Trujillo\\
Av. Juan Pablo s/n, Trujillo, Per\'u.\\
 {\sl  (ctl\_576@yahoo.es)}

\end{center}

\medskip

\medskip
\medskip
\medskip
\medskip

\begin{abstract}
We study the existence of positive solution for the one dimensional Schr\"odinger equation with mixed Lioville-Weyl fractional derivatives
\begin{eqnarray*}\label{Eq00}
_{t}D_{\infty}^{\alpha}({_{-\infty}}D_{t}^{\alpha}u(t)) + V(t) u(t)  = & f(u(t)),\;\;t\in \mathbb{R}\\
u\in H^{\alpha}(\mathbb{R}).\nonumber
\end{eqnarray*}
Furthermore, we analyse radial symmetry property of these solutions. The proof is carried out by using variational methods jointly with comparison and rearrangement argument.

\end{abstract}
\date{}

\setcounter{equation}{0}
\section{ Introduction}
The study of fractional calculus (differentiation and integration of arbitrary order) has emerged as an important and popular field of research. It is mainly due to the extensive application of fractional differential equations in many engineering and scientific disciplines such as physics, chemistry, biology, economics, control theory, signal and image processing, biophysics, blood flow phenomena, aerodynamics, fitting of experimental data, etc., \cite{RH}, \cite{AKHSJT}, \cite{KMBR}, \cite{IP}, \cite{JSOAJTM}, \cite{GZ}. An important characteristic of fractional-order differential operator that distinguishes it from the integer-order differential operator is its nonlocal behavior, that is, the future state of a dynamical system or process involving fractional derivative depends on its current state as well its past states. In other words, differential equations of arbitrary order describe memory and hereditary properties of various materials and processes. This is one of the features that has contributed to the popularity of the subject and has motivated the researchers to focus on fractional order models, which are more realistic and practical than the classical integer-order models.

Very recently, also equations including both left and right fractional derivatives were investigated \cite{OA1, TABS-1, TABS-2, DBJT, MK-0, FJYZ, NNYZ, MTDB, CT-1, CT-2, CT-4, ZZRY}. Equations of this type are known in literature as the fractional Euler-Lagrange equation and are obtained by modifying the principle of least action and applying the rule of fractional integration by parts. Such differential equations mixing both types of derivatives found interesting applications in fractional variational principles, fractional control theory, fractional Lagrangian and Hamiltonian dynamics as well as in the construction industry, see \cite{OA}, \cite{DBSM}, \cite{DB}, \cite{DBSMKT}, \cite{SL}, \cite{SLTB}, \cite{ERKNRHSMDB},  \cite{ES}. Although investigations concerning ordinary and partial fractional differential equations yield many interesting and important results (compare that enclosed in monographies, \cite{AKHSJT}, \cite{KMBR}, \cite{IP}, \cite{SSAKOM}) for equations with operators including fractional derivatives of one type, still the fractional equations with mixed derivatives need further study. This form of fractional operator makes it difficult to find an analytical solution of the considered equation. Some analytical results can be found in papers \cite{OA1}, \cite{DBJT}, \cite{MK-0}, \cite{MK-2} where a fixed point theorem was used. This solution has a complex form, i.e. contains a series of alternately left and right fractional integrals. Using the Mellin transform, Klimek \cite{MK-00} obtained an analytical solution which was represented by a series of special functions. In both cases the analytical results are very difficult for practical calculations.

By other hand it should be noted that critical point theory and variational methods have also turned out to be very effective tools in determining the existence of solutions for integer order differential equations. The idea behind them is trying to find solutions of a given boundary value problem by looking for critical points of a suitable energy functional defined on an appropriate function space \cite{JMMW}, \cite{PR}. In \cite{FJYZ} and \cite{CT-1}, the authors showed that the critical point theory is an effective approach to tackle the existence of solutions for fractional boundary value problem (FBVP) with mixed derivatives. We note that it is not easy to use the critical point theory to study FBVP, since it is often very difficult to establish a suitable space and variational functional for the FBVP.

Inspired by these previous works, in this article we consider the Liouville-Weyl fractional nonlinear Schr\"odinger equation
\begin{eqnarray}\label{Eq00}
_{t}D_{\infty}^{\alpha}({_{-\infty}}D_{t}^{\alpha}u(t)) + V(t) u(t)  = & f(u(t)),\;\;t\in \mathbb{R}\\
u\in H^{\alpha}(\mathbb{R}),\nonumber
\end{eqnarray}
where $\alpha \in (1/2, 1)$, $t\in \mathbb{R}$, $u\in \mathbb{R}$, $f\in C(\mathbb{R})$.

In a recent paper \cite{CT-2}, the author considered nontrivial solution of fractional Hamiltonian systems
\begin{eqnarray}\label{Eq03}
&_{t}D_{\infty}^{\alpha}({_{-\infty}D_{t}^{\alpha}}u(t)) + L(t)u(t)= \nabla W(t,u(t)),
\end{eqnarray} 
where $\alpha \in (1/2,1)$, $L(t)$ is a positive definite $n\times n$ matrix, $W$ is assumed to be superquadratic at infinity and subquadratic  at zero in $u$. It is worth noting that under the assumption $L(t) \to \infty$ as $|t|\to \infty$, the Palais-Smale condition holds and the existence of nontrivial solution of (\ref{Eq03}) follows from the mountain pass theorem. In \cite{CT-3}, Amado, Torres and Zubiaga, considered the potential $W(t,u) = a(t)V(u)$ and assumed that $L$ is uniformly bounded from below and
\begin{equation}\label{infty}
\lim_{|t|\to +\infty} a(t) = 0,
\end{equation}
by using the mountain pass theorem, we derived the existence of nontrivial solution of (\ref{Eq03}). Very recently Nyamoradi and Zhou in \cite{NNYZ} considered 
\begin{eqnarray}\label{bif}
_{t}D_{\infty}^{\alpha}({_{-\infty}}D_{t}^{\alpha}u(t)) + V(t) u(t) -\lambda u(t) = & \mu f(t, u(t)),\;\;t\in \mathbb{R}\\
u\in H^{\alpha}(\mathbb{R}),\nonumber
\end{eqnarray}
under coercivity assumption on $V$ and suitable conditions on $f$, the authors proved the existence of infinitely many solutions for (\ref{bif}) by using the well know critical point theory.

We note that the coercivity assumption for $V$ and $L$ (used on the previous works) is rather strong, and we may wonder if we can relax it. We will show that non-trivial solutions of (\ref{Eq00}) exist under weaker assumptions, but we need a different variational approach

Before continuing, we make precise definition of the notion of solution for the equation
\begin{equation}\label{Eq04}
_{t}D_{\infty}^{\alpha}{_{-\infty}}D_{t}^{\alpha}u(t) + V(t)u(t)  = f(u(t)).
\end{equation}
\begin{Def}\label{def01}
Given $f\in L^{2}(\mathbb{R})$, we say that $u\in H^{\alpha}(\mathbb{R})$ is a weak solution of (\ref{Eq04}) if
$$
\int_{\mathbb{R}}[ {_{-\infty}}D_{t}^{\alpha}u(t) {_{-\infty}}D_{t}^{\alpha}\varphi(t) + V(t)u(t)\varphi(t) ]dt = \int_{\mathbb{R}} f(t,u(t))v(t)dt\;\;\mbox{for all}\;\;v\in H^{\alpha}(\mathbb{R}).
$$
Here $H^{\alpha}(\mathbb{R})$ denotes the fractional Sobolev space (see \S 2).
\end{Def}

Now we state our main assumptions. In order to find solutions of (\ref{Eq00}), we will assume the following general hypotheses.

\begin{itemize}
\item[($f_{0}$)] $f(\xi) \geq  0$ if $\xi \geq 0$ and $f(\xi) = 0$ if $\xi \leq 0$. 
\item[$(f_{1})$] The function $\xi \to \frac{f(\xi)}{\xi}$ is a increasing for $\xi >0$ and $\lim_{\xi \to 0} \frac{f(\xi)}{\xi} = 0$. 

\item[($f_{2}$)] There exists $\theta >2$ such that $\forall t>0$
$$
0 < \theta F(\xi) \leq \xi f(\xi),\;\;\forall \xi,\;\xi \neq 0,\;\;\mbox{where}\;\; F(\xi) = \int_{0}^{\xi} f(\sigma)d\sigma.
$$ 
\item[($f_{3}$)] $\lim_{|\xi|\to \infty} \frac{f(\xi)}{|\xi|^{p_{0}}} = 0$ for some $p_{0} + 1 > \theta$.

\item[$(V_{1})$] $V\in C(\mathbb{R})$ and there is a $V_{0}>0$ such that $V(\xi) \geq V_{0}$ for all $\xi\in \mathbb{R}$.

\item[$(V_{2})$] There is a constant $V_{\infty}$ such that $\liminf_{|x|\to \infty} V(x) \geq V_{\infty}$.
\end{itemize}

Weak solutions to (\ref{Eq00}) are critical points of the functional $I: X^{\alpha} \to \mathbb{R}$ defined by
\begin{equation}\label{X02}
I(u) = \frac{1}{2}\left( \int_{\mathbb{R}} [|{_{-\infty}}D_{t}^{\alpha}u(t)|^2 + V(t)|u(t)|^2] dt\right) - \int_{\mathbb{R}}F(u(t))dt.
\end{equation}
It is standard to check that $I$ is well-defined and of class $C^1$, as a consequence of our assumptions on $f$.

Now we are in a position to state our main existence theorem
\begin{Thm}\label{TM01}
Assume $\frac{1}{2}<\alpha <1$. If $(V_{1}), (V_{2})$ and $(f_{0})-(f_{3})$ hold. Then either $c$ is a critical value of $I$, or $c_{\infty} \leq c$. Moreover this solution satisfies $u(x)\ge 0$ a.e. for all $t\in \mathbb{R}.$
\end{Thm}
Where $c$ and $c_{\infty}$ are the mountain pass critical level associated to $I$ and $I^{\infty}$ respectively (see \S 3).

We prove the existence of weak solution of (\ref{Eq00}) applying the mountain pass theorem \cite{PR} to the functional $I$ defined on $H^{\alpha}(\mathbb{R})$. However,  the direct application of the mountain pass theorem is not possible since Palais-Smale sequences might lose compactness in the whole space $\mathbb{R}$. To overcome this difficulty, we use an argument devised by Rabinowitz in \cite{PR1} comparing the mountain pass critical
value of  $I$ with that of the limiting functional $I^{\infty}$. See the work in \cite{CT-4}, where a similar argument is used. 

As a consequence of Theorem \ref{TM01} we proof the following Theorem
\begin{Thm}\label{TM02}
Assume that
\begin{enumerate}
\item[($V_{3}$)] $\liminf_{|x|\to \infty}V(x) = V_{\infty}$,
\item[($V_{4}$)] $V(t) \leq V_{\infty}$, but $V$ is not identically equal to $V_{\infty}$.
\end{enumerate}
Then $c$ is a critical value for $I$.
\end{Thm}

In our second main theorem we are interested in the symmetric result of weak solution of the equation
\begin{eqnarray}\label{SR00}
_{t}D_{\infty}^{\alpha}({_{-\infty}}D_{t}^{\alpha}u(t)) + V(|t|) u(t)  = & f(u(t)),\;\;t\in \mathbb{R}\\
u\in H^{\alpha}(\mathbb{R}),\nonumber.
\end{eqnarray}
For that purpose we consider that the nonlinearity $f$ satisfies $(f_{0})-(f_{3})$ and the potential V satisfies $(V_{1}), (V_{3})$ and
\begin{itemize}
\item[($V_{5}$)] $V$ is radially symmetric and increasing.
\end{itemize}
Now we state our second Theorem.
\begin{Thm}\label{TM03}
Suppose that $(V_{1)}), (V_{3}), (V_{5})$ and $(f_{0}) - (f_{3})$ hold. Then the mountain pass value is achieved by a radially symmetric function, which  is a  solution of (\ref{SR00}).  
\end{Thm}

To prove this theorem we follow the ideas of Felmer and Torres \cite{PFCT}. We proceed by using rearrangements and variational methods. The idea to prove our result, consists in replacing the path $\gamma$ in the mountain pass setting by its symmetrization $\gamma^{*}: t \in [0,1] \to \gamma (t)^{*}$. Then $u$ would be near of the set $\gamma^{*}([0,1])$. 




The rest of the paper is organized as follows: In $\S$ 2, we describe the Liouville-Weyl fractional calculus and we introduce the fractional space that we use in our work and some proposition are proven which will aid in our analysis. In $\S$ 3 we introduce the Nehari manifold and its properties. In $\S$ 4 we will prove Theorem \ref{TM01} and Theorem \ref{TM02}. In $\S$ 5 we prove Theorem \ref{TM03}.
\section{Preliminary Results}

\subsection{Liouville-Weyl Fractional Calculus}

The Liouville-Weyl fractional integrals of order $0<\alpha < 1$ are defined as
\begin{equation}\label{LWeq01}
_{-\infty}I_{x}^{\alpha}u(x) = \frac{1}{\Gamma (\alpha)} \int_{-\infty}^{x}(x-\xi)^{\alpha - 1}u(\xi)d\xi,
\end{equation}
\begin{equation}\label{LWeq02}
_{x}I_{\infty}^{\alpha}u(x) = \frac{1}{\Gamma (\alpha)} \int_{x}^{\infty}(\xi - x)^{\alpha - 1}u(\xi)d\xi.
\end{equation}
The Liouville-Weyl fractional derivative of order $0<\alpha <1$ are defined as the left-inverse operators of the corresponding Liouville-Weyl fractional integrals
\begin{equation}\label{LWeq03}
_{-\infty}D_{x}^{\alpha}u(x) = \frac{d }{d x} {_{-\infty}}I_{x}^{1-\alpha}u(x),
\end{equation}
\begin{equation}\label{LWeq04}
_{x}D_{\infty}^{\alpha}u(x) = -\frac{d }{d x} {_{x}}I_{\infty}^{1-\alpha}u(x).
\end{equation}

\noindent
We establish the Fourier transform properties of the fractional integral and fractional differential operators. Recall that the Fourier transform $\widehat{u}(w)$ of $u(x)$ is defined by
$$
\widehat{u}(w) = \int_{-\infty}^{\infty} e^{-ix.w}u(x)dx.
$$
Let $u(x)$ be defined on $(-\infty, \infty)$. Then the Fourier transform of the Liouville-Weyl integral and differential operator satisfies
\begin{equation}\label{LWeq06}
\widehat{ _{-\infty}I_{x}^{\alpha}u(x)}(w) = (iw)^{-\alpha}\widehat{u}(w),\;\;\widehat{ _{x}I_{\infty}^{\alpha}u(x)}(w) = (-iw)^{-\alpha}\widehat{u}(w),
\end{equation}
\begin{equation}\label{LWeq08}
\widehat{ _{-\infty}D_{x}^{\alpha}u(x)}(w) = (iw)^{\alpha}\widehat{u}(w), \;\;\widehat{ _{x}D_{\infty}^{\alpha}u(x)}(w) = (-iw)^{\alpha}\widehat{u}(w).
\end{equation}
\subsection{Fractional spaces}

In this section we introduce some fractional derivative space for more details see \cite{CT-2}.

\noindent
Let $\alpha > 0$. Define the semi-norm
$$
|u|_{I_{-\infty}^{\alpha}} = \|_{-\infty}D_{x}^{\alpha}u\|_{L^{2}(\mathbb{R})},
$$
and norm
\begin{equation}\label{FDEeq01}
\|u\|_{I_{-\infty}^{\alpha}} = \left( \|u\|_{L^{2}(\mathbb{R})}^{2} + |u|_{I_{-\infty}^{\alpha}}^{2} \right)^{1/2},
\end{equation}
and let
$$
I_{-\infty}^{\alpha} (\mathbb{R}) = \overline{C_{0}^{\infty}(\mathbb{R})}^{\|.\|_{I_{-\infty}^{\alpha}}}.
$$
Now we define the fractional Sobolev space $H^{\alpha}(\mathbb{R})$ in terms of the Fourier transform. Let $0< \alpha < 1$, let the semi-norm
\begin{equation}\label{FDEeq02}
|u|_{\alpha} = \||w|^{\alpha}\widehat{u}\|_{L^{2}(\mathbb{R})},
\end{equation}
and norm
$$
\|u\|_{\alpha} = \left( \|u\|_{L^{2}(\mathbb{R})}^{2} + |u|_{\alpha}^{2} \right)^{1/2},
$$
and let
$$
H^{\alpha}(\mathbb{R}) = \overline{C_{0}^{\infty}(\mathbb{R})}^{\|.\|_{\alpha}}.
$$

\noindent
We note a function $u\in L^{2}(\mathbb{R})$ belongs to $I_{-\infty}^{\alpha}(\mathbb{R})$ if and only if
\begin{equation}\label{FDEeq03}
|w|^{\alpha}\widehat{u} \in L^{2}(\mathbb{R}).
\end{equation}
Especially
\begin{equation}\label{FDEeq04}
|u|_{I_{-\infty}^{\alpha}} = \||w|^{\alpha}\widehat{u}\|_{L^{2}(\mathbb{R})}.
\end{equation}
Therefore $I_{-\infty}^{\alpha}(\mathbb{R})$ and $H^{\alpha}(\mathbb{R})$ are equivalent with equivalent semi-norm and norm. Analogous to $I_{-\infty}^{\alpha}(\mathbb{R})$ we introduce $I_{\infty}^{\alpha}(\mathbb{R})$. Let the semi-norm
$$
|u|_{I_{\infty}^{\alpha}} = \|_{x}D_{\infty}^{\alpha}u\|_{L^{2}(\mathbb{R})},
$$
and norm
\begin{equation}\label{FDEeq05}
\|u\|_{I_{\infty}^{\alpha}} = \left( \|u\|_{L^{2}(\mathbb{R})}^{2} + |u|_{I_{\infty}^{\alpha}}^{2} \right)^{1/2},
\end{equation}
and let
$$
I_{\infty}^{\alpha}(\mathbb{R}) = \overline{C_{0}^{\infty}(\mathbb{R})}^{\|.\|_{I_{\infty}^{\alpha}}}.
$$
Moreover $I_{-\infty}^{\alpha}(\mathbb{R})$ and $I_{\infty}^{\alpha}(\mathbb{R})$ are equivalent, with equivalent semi-norm and norm \cite{CT-2}.
\begin{Thm}\label{FDEtm01}
\cite{CT-2} If $\alpha > \frac{1}{2}$, then $H^{\alpha}(\mathbb{R}) \subset C(\mathbb{R})$ and there is a constant $C=C_{\alpha}$ such that
\begin{equation}\label{FDEeq06}
\|u\|_{\infty} \leq C \|u\|_{\alpha}.
\end{equation}
\end{Thm}
\begin{Remark}\label{FDEnta01}
If $u\in H^{\alpha}(\mathbb{R})$, then $u\in L^{q}(\mathbb{R})$ for all $q\in [2,\infty]$, since
$$
\int_{\mathbb{R}} |u(x)|^{q}dx \leq \|u\|_{\infty}^{q-2}\|u\|_{L^{2}(\mathbb{R})}^{2}.
$$
\end{Remark}

Let 
$$
X^{\alpha} = \left\{ u\in H^{\alpha}(\mathbb{R})/\;\; \int_{\mathbb{R}} (|{_{-\infty}}D_{t}^{\alpha}u(t)|^2 + V(t)|u(t)|^{2})dt < \infty \right\}.
$$
The space $X^{\alpha}$ is a reflexive and separable Hilbert space with the inner product
\begin{equation}\label{X01}
\langle u,v \rangle_{X^\alpha} = \int_{\mathbb{R}} ({_{-\infty}}D_{t}^{\alpha}u(t) {_{-\infty}}D_{t}^{\alpha}v(t) + V(t)u(t)v(t))dt
\end{equation}
and the corresponding norm
\begin{equation}\label{X02}
\|u\|_{X^{\alpha}}^{2} = \langle u,u \rangle_{X^{\alpha}}.
\end{equation}
Similarly to the proofs of Lemma 2.1 in \cite{CT-2} we can get the following lemma.
\begin{Lem}\label{Xlm01}
Suppose $V(t)$ satisfies $(V_{1})$. Then the space $X^{\alpha}$ is continuously embedded in $H^{\alpha}(\mathbb{R})$.
\end{Lem}

One major tool in variational methods is the following version of the concentration compactness principle, originally proved by P.L. Lions.
\begin{Lem}\label{XCC}
\cite{CT-4} Let $r>0$ and $q\geq 2$. Let $(u_{n}) \in H^{\alpha}(\mathbb{R})$ be bounded. If
\begin{equation}\label{FDECeq01}
\lim_{n\to \infty} \sup_{y\in \mathbb{R}} \int_{y-r}^{y+r} |u_{n}(t)|^{q}dt \to 0,
\end{equation}
then $u_{n} \to 0$ in $L^{p}(\mathbb{R})$ for any $p>2$.
\end{Lem} 

\section{The Nehari Manifold and Qualitative Properties of Ground Sate Levels}

In this section we introduce the Nehari manifold associated to $I$ as 
$$
\mathcal{N} = \left\{ u\in X^{\alpha}\setminus \{0\}/\;\; I'(u)u=0 \right\},
$$
and we observe that all non trivial solutions of (\ref{Eq00}) belong to $\mathcal{N}$. Next, from $(f_{1})$ and $(f_{3})$) it is standard to prove that,  for any $\epsilon >0$, there exists $C_{\epsilon}$ such that
\begin{equation}\label{limeq05}
|f(\xi)| \leq \epsilon |\xi| + C_{\epsilon} |\xi|^{p_{0}},\;\;\forall t \in \mathbb{R},
\end{equation}
and consequently 
\begin{equation}\label{limeq06}
|F(\xi)| \leq \frac{\epsilon}{2} |\xi|^{2} + \frac{C_{\epsilon}}{p_{0} + 1} |\xi|^{p_{0}+1},\;\;\forall t \in \mathbb{R}.
\end{equation}
We start our analysis with

\begin{Lem}\label{limlm01}
Assume the hypotheses ($f_{0}$)-($f_{3}$) hold. For any $u\in X^{\alpha}(\mathbb{R})\setminus \{0\}$, there is  a unique $\sigma_{u} = t(u) > 0$ such that $\sigma_{u}u \in \mathcal{N}$ and we have
$$
I(\sigma_{u}u) = \max_{\sigma \geq 0} I(\sigma u).
$$
\end{Lem}

\noindent
{\bf Proof.} Let $u\in X^{\alpha}(\mathbb{R})\setminus \{0\}$ and consider the function
$
\psi  :  \mathbb{R}^{+} \to \mathbb{R}$ defined as 
     \begin{eqnarray*} \psi (\sigma) = I(\sigma u) = \frac{\sigma ^{2}}{2}\|u\|_{X^\alpha}^{2} - \int_{\mathbb{R}}F(\sigma u)dt.
\end{eqnarray*}
Then, by (\ref{limeq06}) we have
\begin{eqnarray*}
\int_{\mathbb{R}}F(u)dt & \leq & \frac{C\epsilon}{2}\|u\|_{X^\alpha}^{2} + \frac{CC_{\epsilon}}{p_{0} + 1}\|u\|_{X^\alpha}^{p_{0}+1}.
\end{eqnarray*}
This implies that
$
\psi (\sigma) > 0,\;\;\mbox{for}\;\;\sigma \;\;\mbox{small}. 
$
On the other hand, by ($f_{2}$) there exists $A>0$ such that $F(\xi) \geq A|\xi|^{\theta},\;\;\forall |\xi|>1$. So
\begin{eqnarray}\label{limeq07}
I(\sigma u) & \leq & \frac{\sigma^{2}}{2}\|u\|_{X^\alpha}^{2} - A\sigma^{\theta}\int_{\mathbb{R}}|u(t)|^{\theta}dt,
\end{eqnarray}
and since $\theta > 2$, we see that $\psi (\sigma) < 0$ for $\sigma$ large. By ($f_{0}$), $\psi (0) = 0$, therefore there is $\sigma_{u} = \sigma(u) > 0$ such that
$$
\psi (\sigma_{u}) = \max_{\sigma\geq 0} \psi (\sigma) = \max_{\sigma \geq 0} I(\sigma u) = I(\sigma_{u}u).
$$
We see that $\psi'(\sigma) = 0$ is equivalent to
\begin{eqnarray*}
\|u\|_{X^\alpha}^{2} = \int_{\mathbb{R}}\frac{f(\sigma u)u}{\sigma }dx,
\end{eqnarray*}
from where, using ($f_{1}$) we prove that   there is a unique $\sigma_{u} >0$ such that
$
\sigma_{u}u \in \mathcal{N}.
$
$\Box$

In the sequel, we will need to estimate the behavior of $I$ on $\mathcal{N}$. The following identities will be useful.

\begin{Lem}\label{limlm02}
Define
$$
c^{*} = \inf_{u\in X^{\alpha}\setminus \{0\}} \max_{\sigma \geq 0} I(\sigma u)
$$
then 
$$
c^{*} = c = \inf_{u\in \mathcal{N}}I(u).
$$
\end{Lem}

\noindent
{\bf Proof.} The proof is rather standard. The identity $c^{*} = \inf_{\mathcal{N}}I$ is a trivial consequence of the previous Lemma. To prove $c = \inf_{\mathcal{N}}I$, we fix an arbitrary $u\in \mathcal{N}$ and define a path $\gamma_{u}$ as follows: $\gamma_{u}(\sigma) = \sigma (\sigma_{u}u)$, where $I(\sigma_{u}u)<0$. Since $\gamma_{u} \in \Gamma$, $c\leq \inf_{\mathcal{N}}I$. On the other hand, if $\gamma \in \Gamma$, then $\gamma (\sigma) \in \mathcal{N}$ for some $\sigma \in (0,1)$. Indeed, if $I'(\gamma (\sigma))\gamma (\sigma) >0$, then $I(\gamma (\sigma)) \geq 0$ for every $\sigma$, and this contradicts the fact that $I(\gamma (1))<0$. $\Box$   

In the rest of this section, we will study some qualitative properties of the level $c$ as a function of the potential $V$. For this reason, we introduce the provisional notation $c_{V}$ for $c$.   

\begin{Prop}\label{limprop01}
Let $f$ satisfy $(f_{0})-(f_{3})$ and $V$, $\overline{V}$, satisfy $(V_{1})$. If $V \geq \overline{V}$, then $c_{V} \geq c_{\overline{V}}$
\end{Prop}

\noindent
{\bf Proof.} If $V \geq \overline{V}$, then if $\overline{I}$ is the functional associated with $\overline{V}$, 
\begin{equation}\label{Q01}
I(u) \geq \overline{I}(u)
\end{equation}
for all $u \in X^{\alpha}$. Let $\overline{\Gamma}$ be the analogue of $\Gamma$ for $\overline{I}$. Then $\gamma \in \Gamma$ implies $\gamma \in \overline{\Gamma}$ and by (\ref{Q01}),
\begin{equation}\label{Q02}
\max_{\sigma \in [0,1]} I(\gamma (\sigma)) \geq \max_{\sigma \in [0,1]} \overline{I}(\gamma (\sigma)).
\end{equation}
Consequently
$$
c\geq \inf_{\gamma \in \Gamma} \max_{\sigma \in [0,1]} \overline{I}(\gamma (\sigma)) \geq \inf_{\gamma \in \overline{\Gamma}} \max_{\sigma \in [0,1]} \overline{I}(\gamma (\sigma)) = c_{\overline{V}}.
$$
$\Box$

This monotonicity is the key to prove the continuity of $c_{V}$ with respect to $V$.

\begin{Prop}\label{limprop02}
Let $f$ satisfy $(f_{0})-(f_{3})$ and $V, V_{n}$ satisfy $(V_{1})$ for $n\in \mathbb{N}$. If $V_{n} \to V$ uniformly, then $c_{V_{m}} \to c_{V}$.
\end{Prop}  

\noindent
{\bf Proof.} Let $\epsilon >0$. Then for large $n$,
$$
V + \epsilon \geq V + |V_{n} - V| \geq V \geq V - |V_{n} - V| \geq V-\epsilon
$$
By the monotonicity of $c_{V}$, it is enough to prove the weaker result
$$
\lim_{\epsilon \to 0} c_{(V + \epsilon)} = c_{V} .
$$
Put, to make notation lighter, $c_{\epsilon} = c_{(V + \epsilon)}$. By  Proposition \ref{limprop01},
$$
\lim_{\epsilon \to 0^{-}} c_{\epsilon} = c_{-} \leq c_{V} = c_{0}.
$$
Suppose that 
\begin{equation}\label{cc}
c_{-} < c_{0},
\end{equation} 
and consider the functional
\begin{equation}\label{Q03}
I_{\epsilon}(u) = \frac{1}{2} \int_{\mathbb{R}}[ |{_{-\infty}}D_{t}^{\alpha}u(t)|^{2} + (V(t)+\epsilon)|u(t)|^{2}]dt - \int_{\mathbb{R}} F(u(t))dt
\end{equation}
Let $\epsilon_{k} \to 0^{-}$ as $k\to \infty$ and $\delta_{n} \to 0^{+}$ as $n\to \infty$. For each such $k$, by Lemma \ref{limlm02}, there is a sequence $u_{kn} \in X^{\alpha}$ such that $\|u_{kn}\|_{X^{\alpha}} = 1$ and 
$$
\max_{\sigma \geq 0} I_{\epsilon_{k}}(\sigma u_{kn}) \leq c_{\epsilon_{k}} + \delta_{n}.
$$
To each $u_{kn}$, we associated a path $\gamma_{kn}$ such that
$$
\max_{0\leq t \leq 1} I_{\epsilon_{k}}(\gamma_{kn}(t)) = \max_{\sigma\geq 0}I_{\epsilon_{k}}I(\sigma u_{kn}).
$$
By a Theorem of Mawhin-Willem - Theorem 4.3 of \cite{JMMW}, there are sequences $\{w_{kn}\} \in X^{\alpha}$ and $t_{kn}\in [0,1]$ such that
\begin{eqnarray*}
&& \|w_{kn} - \gamma_{kn}(t_{kn})\|_{X^{\alpha}} \leq \sqrt{\delta_{n}}\\
&&I_{\epsilon_{k}}(w_{kn}) \in (c_{\epsilon_{k}} - \delta_{n}, c_{\epsilon_{k}})\\
&& \|I'_{\epsilon_{k}}(w_{kn})\| \leq \sqrt{\delta_{n}}
\end{eqnarray*}
By (\ref{Q03})
$$
I_{\epsilon_{k}} (u) = I(u) + \frac{\epsilon_{k}}{2} \int_{\mathbb{R}}|u(t)|^{2}dt,\;\;\forall u\in X^{\alpha}.
$$
Taking $n=k$ above, set $u_{k} = u_{kk}$ and $w_{k} = w_{kk}$. Then
\begin{eqnarray}\label{Q04}
c_{0} &\leq& \max_{\sigma \geq 0}I(\sigma u_{k}) = I(\sigma_{u_{k}}u_{k})\nonumber\\
& = & I_{\epsilon_{k}}(\sigma_{u_{k}}u_{k}) - \frac{\epsilon_{k}}{2}\sigma_{u_{k}}^2\int_{\mathbb{R}}|u_{k}(t)|^2dt\nonumber\\
&\leq& \max_{\sigma \geq 0} I_{\epsilon_{k}}(\sigma_{u_k}u_k) - \frac{\epsilon_{k}}{2}\sigma_{u_k}^2 \|u_{k}\|_{L^2}^{2}\nonumber\\
&\leq& c_{\epsilon_{k}} + \delta_{k} - \frac{\epsilon_{k}}{2}\sigma_{u_k}^2 \|u_{k}\|_{L^2}^{2}\nonumber\\
&\leq& c_{-} + \delta_{k}  - \frac{\epsilon_{k}}{2} \sigma_{u_k}^2 \|u_{k}\|_{L^2}^{2}.
\end{eqnarray}
Since $\|u_{k}\|_{X^{\alpha}} = 1$, by the continuous embedding of $X^{\alpha}$ in $L^{2}(\mathbb{R})$, There is a constant $M_{1}>0$ such that $\sup_{k}\|u_{k}\|_{L^{2}} \leq M_{1}$. Hence if $\{\sigma_{u_k}\}$ is bounded independently of $k$, by choosing $\epsilon_{k}$ small enough, (\ref{Q04}) is contrary to (\ref{cc}). Recalling the definition of $\sigma_{u_k}$ and the normalization of $(u_{k})$,
\begin{equation}\label{Q05}
\sigma_{u_k}^2 = \int_{\mathbb{R}}\sigma_{u_k}u_{k}(t)f(\sigma_{u_k}u_{k}(t))dt.
\end{equation}   
If along a subsequence $\sigma_{u_k} \leq 1$, we are through. Otherwise for large $k$, $\sigma_{u_k} >1$ and by ($f_{2}$),
$$
\sigma_{u_k}^{2} \geq \theta \int_{\mathbb{R}} F(\sigma_{u_k}u_{k}(t))dt \geq \theta \sigma_{u_k}^{\theta} \int_{\mathbb{R}} F(u_{k}(t))dt,
$$
so
\begin{equation}\label{Q06}
\sigma_{u_k}^{\theta -2} \leq \left( \theta \int_{\mathbb{R}} F(u_{k}(t))dt \right)^{-1}.
\end{equation}
Since there is no upper bound for $\sigma_{u_k}$, the denominator must approach zero as $k\to +\infty$. But this is impossible. Indeed, since $\gamma_{kn}(\sigma) = (\sigma_{u_{kn}}u_{kn})\sigma$, then, for $n=k$ we have
$$
\gamma_{k}(\sigma_{k}) = (\sigma_{u_{kk}}u_{kk})\sigma_{kk} = \xi_{k}u_{k}
$$ 
The properties of $w_{k}$ imply now that
$$
\|w_{k} - \xi_{k}u_{k}\|_{X^{\alpha}} \leq \sqrt{\delta_{k}}.
$$
Since $\{w_{k}\}$ is bounded, there is a constant $M_{2}>0$ such that
$$
\xi_{k} \leq \sqrt{\delta_{k}} + \|w_{k}\|_{X^{\alpha}} \leq M_{2}.
$$
For any ball $B(y,r)$, we have
\begin{eqnarray*}
\|u_{k}\|_{L^{2}(B(y,r))} &\geq& M_{2}^{-1}\|\xi_{k}u_{k}\|_{L^{2}(B(y,r))}\\
&\geq& M_{2}^{-1} (\|w_{k}\|_{L^{2}(B(y,r))} - \|w_{k} - \xi_{k}u_{k}\|_{L^{2}(B(y,r))})\\
&\geq& M_{2}^{-1} (\|w_{k}\|_{L^{2}(B(y,r))} - M_{3}\sqrt{\delta_{k}})
\end{eqnarray*}
By Lemma \ref{XCC}, there are a sequence of points $\{y_{k}\}$ and numbers $\beta, R>0$ such that
$$
\liminf_{k\to +\infty} \int_{B(y_{k}, R)}|w_{k}(t)|^{2} \geq \beta.
$$
Hence, for $k$ large enough
\begin{equation}\label{Q07}
\|u\|_{L^{2}(B(y_{k}, R))} \geq M_{2}^{-1}\sqrt{\frac{\beta}{2}}.
\end{equation}
Recall that we want to prove that 
$$
\int_{\mathbb{R}} F(u_{k}(t))dt \to 0 
$$
is impossible. From ($f_{3}$), given $\lambda >0$, there exists $A_{\lambda} >0$ such that
$$
|s|^{2} \leq \lambda + A_{\lambda} F(s),\;\;\forall s\in \mathbb{R}.
$$
Consequently
$$
\int_{B(y_{k}, R)} |u_{k}(t)|^2dt \leq \lambda + A_{\lambda}\int_{B(y_{k}, R)}F(u_{k}(t))dt.
$$
If 
$$\int_{\mathbb{R}}F(u_{k}(t))dt \to 0,$$
then 
$$
\int_{B(y_{k}, R)} |u_{k}(t)|^{2}dt \to 0,
$$
contrary to (\ref{Q07}).

We have finally proved that 
$$
\lim_{\epsilon \to 0^-} c_{\epsilon} = c_{0}.
$$
To complete the proof, assume by contradiction that
$$
c_{0} < c_{+} = \lim_{\epsilon \to 0^{+}} c_{\epsilon}.
$$
Let $\delta_{k}$ be as before; again, there is a sequence $\{u_{k}\}$ in $X^{\alpha}$ such that $\|u_{k}\|_{X^{\alpha}} = 1$ and
$$
\max_{\sigma\geq 0}I(\sigma u_{k}) = c_{0} + \delta_{k}
$$ 
Choose $w_{k} = w_{kk}$ as above. For each $\epsilon \geq 0$ and $u\in X^{\alpha} \setminus \{0\}$, let $\sigma_{u}^{\epsilon}$ play the role for $I_{\epsilon}$ that $\sigma_{u}$ does for $I$. Hence
\begin{eqnarray*}
c^+ &\leq &c_{\epsilon} \leq \max_{\sigma \geq 0} I_{\epsilon}(\sigma u_{k}) = I_{\epsilon}(\sigma_{u_k}^{\epsilon}u_{k})\\
& = & I(\sigma_{u_k}^{\epsilon}u_{k}) + \epsilon (\sigma_{u_k}^{\epsilon})^{2} \|u_{k}\|_{L^{2}}^{2}\\
&\leq& c_{0} + \delta_{k} + \epsilon (\sigma_{u_k}^{\epsilon})^{2}\|u_{k}\|_{L^2}^{2}.
\end{eqnarray*}
As above, either $\sigma_{u_k}^{\epsilon} \leq 1$ or
$$
(\sigma_{u_k}^{\epsilon})^{\theta - 2} \leq \frac{\int_{\mathbb{R}}[|{_{-\infty}}D_{t}^{\alpha}u(t)|^2 + (V(t) + \epsilon)|u_{k}(t)|^2] dt}{\theta \int_{\mathbb{R}}F(u_{k}(t))dt}.
$$
In any case, we can conclude as earlier that $\{\gamma_{\epsilon}(u_{k})\}$ is a bounded sequence, and $c_{0} < c^+ = \lim_{\epsilon \to 0^+}c_{\epsilon}$ cannot hold. This completes the proof. $\Box$

\section{Existence Result}

In this section some existence result will be established for (\ref{Eq00}). They are based in part on comparison arguments. Thus let $V_{\infty}$ given by ($V_{2}$) and set 
\begin{equation}\label{er01}
I^{\infty}(u) = \frac{1}{2}\int_{\mathbb{R}} [|{_{-\infty}}D_{t}^{\alpha}u(t)|^2 + V_{\infty}|u(t)|^{2}]dt - \int_{\mathbb{R}}F(u(t))dt. 
\end{equation}
This functional is of class $C^{1}$ and has the mountain pass geometry (see \cite{CT-4}); hence, we can set
$$
\Gamma^{\infty} = \{\gamma \in C([0,1], X^{\alpha})/\;\;\gamma (0) = 0, I^{\infty}(\gamma (1)) <0\}
$$
and
$$
c_{\infty} = \inf_{\gamma \in \Gamma^{\infty}} \max_{0\leq \sigma \leq 1}I^{\infty}(\gamma (\sigma)).
$$

\noindent
{\bf Proof of Theorem \ref{TM01}.} We first prove the theorem under the stronger assumption
\begin{equation}\label{er02}
\liminf_{|x|\to +\infty}V(x) \geq V_{\infty}.
\end{equation}
The proof follows in part arguments from Proposition \ref{limprop02}. As earlier, the different characterization of the level $c$ provides a sequence $\{u_{n}\}$ in $X^{\alpha}$ such that $\|u_{n}\|_{X^{\alpha}} = 1$ and
\begin{equation}\label{clim}
\max_{\sigma \geq 0} I(\sigma u_{n}) \to c,
\end{equation}
as $n\to \infty$. Attach a path $\gamma_{n} \in \Gamma$ to each $u_{n}$ is such a way that
$$
\max_{0\leq t\leq 1} I(\gamma_{n}(t)) = \max_{\sigma \geq 0} I(\sigma u_{n}).
$$
Once again, we can find sequence $\{w_{n}\} $ in $X^{\alpha}$, $\epsilon_{n} \to 0$ and $t_{n}\in [0,1]$ such that
\begin{eqnarray*}
&& \|w_{n} - \gamma_{n}(t_{n})\|_{X^{\alpha}} \leq \sqrt{\epsilon_{n}},\\
&& I(w_{n}) \in (c-\epsilon_{n}, c),\\
&&\|I'(w_{n})\| \leq \sqrt{\epsilon_{n}}.
\end{eqnarray*}
It follows easily that $\{w_{n}\}$ is bounded, and assume that, up to subsequences, it converges weakly in $X^{\alpha}$ to some $w$ and strongly in $L^{p}_{loc}(\mathbb{R})$, for any $p\in [2,\infty)$. Then $w$ weakly solves the limiting equation
\begin{equation}\label{er03}
{_{t}}D_{\infty}^{\alpha}{_{-\infty}}D_{t}^{\alpha}w + V_{\infty}w = f(w).
\end{equation}
Lemma \ref{XCC} implies the existence of a sequence of points $\{y_{n}\} \in \mathbb{R}$ and of constant $\beta >0$ and $R>0$ such that
$$
\liminf_{n\to +\infty} \int_{B(y_{n}, R)} |w_{n}(t)|^2dt >\beta.
$$
If the sequence $\{y_{n}\}$ is bounded, then $w\neq 0$ and the local compactness of the Sobolev embedding tells us that, for every $\rho >0$,
\begin{eqnarray}\label{er04}
I(w_{n}) - \frac{1}{2}I'(w_{n})w_{n} &=& \int_{\mathbb{R}}\left( \frac{1}{2}f(w_{n}(t))w_{n}(t) - F(w_{n}(t)) \right)dt\nonumber \\
&\geq& \int_{B(0,\rho)} \left(\frac{1}{2}f(w_{n}(t))w_{n}(t)\!\!  - F(w_{n}(t))\right)dt\nonumber\\
&\to&  \int_{B(0,\rho)}\left( \frac{1}{2}f(w(t))w(t) - F(w(t)) \right)dt
\end{eqnarray}
Since the left hand of (\ref{er04}) approaches $c$ as $n\to \infty$ and $\rho$ is arbitrary,
$$
c\geq \int_{\mathbb{R}}\left(\frac{1}{2}f(w(t))w(t) - F(w(t)) \right)dt.
$$
But the right-hand side of this relation coincides with $I^{\infty}(w)$, since $w$ solves (\ref{er03}), and therefore
$$
c\geq c_{\infty}.
$$
If, on the other hand, $\{y_{n}\}$ is unbounded, and we may even assume that $y_{n} \to +\infty$, then, for every $\xi >0$ and $\rho >0$,
\begin{eqnarray*}
\max_{\sigma \geq 0} I(\sigma u_{n}) &\geq&I(\xi u_{n}) =  I^{\infty}(\xi u_{n}) + \frac{1}{2}\int_{B(0,\rho)} (V(t) - V_{\infty})|\xi u_{n}(t)|^2 dt\\
&& + \frac{1}{2}\int_{\mathbb{R} \setminus B(0,\rho)}(V(t) - V_{\infty})|\xi u_{n}(t)|^2dt. 
\end{eqnarray*} 
Thanks to assumption (\ref{er02}) we may choose $\rho >0$ so that $V(t) \geq V_{\infty}$ whenever $|t| \geq \rho$. Thus
$$
\max_{\sigma \geq 0} I(\sigma u_{n}) \geq I^{\infty}(\xi u_{n}) + \frac{1}{2} \int_{B(0,\rho)} (V(t) - V_{\infty})|\xi u_{n}(t)|^2 dt.
$$
Specialize now $\xi = \sigma_{u_{n}}^{\infty}$, where $\sigma_{u_{n}}^{\infty}$ is the unique positive number such that $\sigma_{u_{n}}^{\infty}u_{n}$ belongs to the Nehari manifold of $I^{\infty}$. As such
$$
I^{\infty}(\sigma_{u_{n}}^{\infty}u_{n}) = \max_{\sigma \geq 0}I^{\infty}(\sigma u_{n})
$$
and 
$$
\max_{\sigma \geq 0} I(\sigma u_{n}) \geq c_{\infty} + \frac{1}{2}\int_{B(0,\rho)}(V(x) - V_{\infty})|\sigma_{u_{n}}^{\infty}u_{n}(t)|^2dt.
$$
The arguments of the proof of Proposition \ref{limprop02} show $\{\sigma_{u_{n}}^{\infty}\}$ is bounded. Suppose there is a $\Lambda >0$ such that
\begin{equation}\label{er05}
\int_{B(0,\rho)}|u_{n}(t)|^2dt \geq \Lambda
\end{equation}
As in Proposition \ref{limprop02}, $\gamma_{n}(t_{n}) = \xi_{n}u_{n}$ and by the properties of $w_{n}$ 
\begin{equation}\label{er06}
\|w_{n} - \xi_{n}u_{n}\|_{X^{\alpha}} \leq \sqrt{\epsilon_{n}}
\end{equation}
Therefore
\begin{eqnarray}\label{er07}
\|w_{n}\|_{L^{2}(B(0,\rho))} & \geq & \|\xi_{n}u_{n}\|_{L^{2}(B(0,\rho))} - \|w_{n} - \xi_{n}u_{n}\|_{L^{2}(B(0,\rho))}.
\end{eqnarray}
By (\ref{er06}), the $w_{n}$ term on the right hand side of (\ref{er07}) tends to $0$ as $n\to \infty$.
If $\xi_{n} \to$ along a subsequence, $\xi_{n}u_{n} \to 0$ and $I(\xi_{n}u_{n})\to 0$, contrary to (\ref{clim}). Hence $\{\xi_{n}\}$ has a positive lower bound and (\ref{er07}) shows there is a $\Lambda_{1}$ such that
\begin{equation}\label{er08}
\|w_{n}\|_{L^{2}(B(0,\rho))} \geq \Lambda_{1}.
\end{equation}
Therefore as for the case of bounded $\{y_{n}\}$, $w_{n}$ converges weakly in $X^{\alpha}$, along a subsequence, to $w$ a solution of (\ref{Eq00}) with $I(w) = c$.

To complete the proof, we must show that (\ref{er05}) is true. If not, along a subsequence, 
$$
\|u_{n}\|_{L^{2}(B(0,\rho))} \to 0.
$$ 
But then
\begin{eqnarray*}
c + o(1) & =& \max_{\sigma \geq 0}I(\sigma u_{n}) \geq c_{\infty} + \frac{1}{2}\int_{B(0,\rho)} (V(t) - V_{\infty})|\sigma_{u_{n}}^{\infty}u_{n}(t)|^2dt\\
& = & c_{\infty} + o(1),
\end{eqnarray*}
i.e., $c \geq c_{\infty}$. The proof is complete under the stronger assumption (\ref{er02}).

Suppose now that 
$$
\liminf_{|x| \to +\infty}V(x) = V_{\infty}.
$$
Pick $\epsilon >0$ so that
$$
\liminf_{|x| \to +\infty}V(x) > V_{\infty} - \epsilon.
$$
We can apply the previous proof to the potential $V_{\epsilon} = V- \epsilon$: hence, either (i) $c$ is a critical value of $I$ or (ii) $c \geq c_{\infty}^{\epsilon}$, where $c_{\infty}^{\epsilon}$ is the valued obtained above, on replacing $I^{\infty}$, $\Gamma^{\infty}$ by $I_{\epsilon}^{\infty}$, $\Gamma_{\epsilon}^{\infty}$ in the obvious fashion. Suppose that (i) does not hold. Letting $\epsilon \to 0$ and using Proposition \ref{limprop02} then yields $c\geq c_{\infty}$ for this case. The proof in complete. 

Having the existence of a critical point $u$ of $I$ in $H^{\alpha}(\mathbb{R})$, we just have to prove that $u \geq 0$ a.e. For this fact, we recall that
$$
\int_{\mathbb{R}} |w|^{2\alpha} \widehat{u} \widehat{\varphi}dw = C \int_{\mathbb{R}}\int_{\mathbb{R}} \frac{[u(x) + u(y)][\varphi (x) - \varphi (y)]}{|x-y|^{1+2\alpha}}dxdy,
$$
for all $\varphi \in H^{\alpha}(\mathbb{R})$, see \cite{MW}. Testing with $u_{-}:= \max \{-u, 0\}$, by the positive of $f(u(t))$ we obtain
$$
\int_{\mathbb{R}} |w|^{2\alpha} \widehat{u} \widehat{\varphi}dw = \int_{\mathbb{R}} V(t)u_{-}^{2}dt.
$$
But this cannot occur for $u_{-} \not \equiv 0$, because
\begin{eqnarray*}
\int_{\mathbb{R}} |w|^{2\alpha} \widehat{u} \widehat{\varphi}dw &=& C \int_{\{u<0\}}\int_{\{u>0\}} \frac{[u(x) + u(y)]u_{-}(x)}{|x-y|^{1+2\alpha}}dxdy\\
&& +C \int_{\{u>0\}}\int_{\{u<0\}} \frac{[u(x) + u(y)]u_{-}(y)}{|x-y|^{1+2\alpha}}dxdy\\
&& + C\int_{\{u<0\}}\int_{\{u<0\}} \frac{[u(x) + u(y)][u_{-} (x) - u_{-} (y)]}{|x-y|^{1+2\alpha}}dxdy.
\end{eqnarray*}
The last term can be written as
$$
-C\int_{\{u<0\}}\int_{\{u<0\}} \frac{|u_{-}(x)] + u_{-}(y)|^{2}}{|x-y|^{1+2\alpha}}dxdy,
$$
which is strictly negative unless $u_{-} \equiv 0$ a.e. The other two terms are also negative, hence, $u_{-} \equiv 0$ and the conclusion follows. 
$\Box$ 


\noindent
{\bf Proof of Theorem \ref{TM02}.} If $c$ is not a critical value of $I$, by Theorem \ref{TM01}, $c\geq c_{\infty}$. Let $w$ be any critical point of $I^{\infty}$ corresponding to $c_{\infty}$ as given by Theorem 3.1 of \cite{CT-4}. Then
\begin{equation}\label{er09}
c_{\infty} = I^{\infty}(w) = \max_{\sigma \geq 0} I^{\infty}(\sigma w).
\end{equation}
For any $\sigma >0$,
\begin{equation}\label{er10}
I^{\infty}(\sigma w) = I(\sigma w) + \frac{1}{2} \int_{\mathbb{R}} (V_{\infty} - V(t))|\sigma w(t)|^2dt.
\end{equation}
Choose $\sigma = \sigma_{u}$. By (\ref{er09})-(\ref{er10}) and ($V_{4}$), 
\begin{eqnarray*}
c_{\infty} &\geq& I(\sigma_{w}w) + \frac{1}{2}\int_{\mathbb{R}} (V_{\infty} - V(t)) |\sigma_{u}w(t)|^2dt\\
&\geq& c + \frac{1}{2} \int_{\mathbb{R}} (V_{\infty} - V(t))|\sigma_{u}w(t)|^2dt >c
\end{eqnarray*}
contrary to Theorem \ref{TM01}. 
$\Box$

\section{Symmetry Results}

\subsection{Tools: Symmetry Rearrangement}

In this section first we recall some facts regarding rearrangement of sets and functions. Then we present a new regional Riesz and Polya-Szeg\"o inequality when the range of scope determined is a radially symmetric function.   

Let $A\subset \mathbb{R}$ be a Lebesgue measurable set and denote the measure of $A$ by $|A|$. Define the symmetrization $A^{*}$ of $A$ to be the closed ball centered at the origin such with the same measure as $A$. Thus 
$$
A^{*} := [-\frac{|A|}{2}, \frac{|A|}{2}]
$$

Let $u:\mathbb{R} \to \mathbb{R}$ a Borel measurable function, then $u$ is said to vanish at infinity if
$$
|\{x: |u(x)|>t\}|< \infty\;\;\mbox{for all}\;\;t>0
$$

The symmetric decreasing rearrangement of a characteristic function $\chi_{A}$ is defined as
$$
\chi_{A}^{*}:= \chi_{A^{*}}
$$

We now use that any non negative function can be expressed as an integral of the characteristic functions of the sets $\{u\geq t\}$ (which is a standard abbreviation for $\{x: u(x)\geq t\}$) as follows
\begin{equation}\label{SReq1}
u(x) = \int_{0}^{u(x)}1dt = \int_{0}^{\infty} \chi_{\{u\geq t\}}(x)dt.
\end{equation}
Note that this, along with Fubini's theorem, implies
\begin{eqnarray*}
\int_{\mathbb{R}^{n}} u(x)dx & = & \int_{\mathbb{R}} \int_{0}^{\infty} \chi_{\{ u \geq t \}}(x)dtdx\\
& = & \int_{0}^{\infty} |\{x: u(x) \geq t\}|dt.
\end{eqnarray*}
Now if $u:\mathbb{R}^{n} \to \mathbb{R}$ is a Borel measurable function vanishing at infinity we define
\begin{equation}\label{SReq2}
u^{*}(x) = \int_{0}^{\infty} \chi_{\{|u|\geq t\}}^{*}(x)dt
\end{equation}

The rearrangement $u^{*}$ has a number of properties, see \cite{ELML}:
\begin{itemize}
\item[(i)] $u^{*}$ is nonnegative.
\item[(ii)] $u^{*}$ is radially symmetric and nonincreasing, i.e:
\begin{eqnarray*}
|x| \leq |y| \;\;\mbox{implies}\;\;u^{*}(y) \leq u^{*}(x)
\end{eqnarray*}
\item[(iii)] $u^{*}$ is a lower semicontinuos function.
\item[(iv)] The level sets of $u^{*}$ are simply the rearrangement of the level set of $u$, i.e
 $$
 \{x:u^{*}(x) > t\} = \{x:|u(x)|>t\}^{*}.
 $$
 an important consequence of this is the equimeasurability  of the function $u$ and $u^{*}$, i.e
\begin{eqnarray*}
|\{u^{*} > t\}| = |\{|u| > t\}|\;\;\mbox{for all}\;\;t>0.
\end{eqnarray*}
\item[(v)] For any positive monotone function $\phi$, we have
$$
\int_{\mathbb{R}^{n}} \phi (|u(x)|)dx = \int_{\mathbb{R}^{n}} \phi (u^{*}(x))dx.
$$
In particular, $u^{*}\in L^{p}(\mathbb{R})$ if and only if $u\in L^{p}(\mathbb{R})$ and
$$
\|u\|_{L^{p}} = \|u^{*}\|_{L^{p}}
$$
\item[(vi)] Let $V(|x|) \geq 0$ be a spherically symmetric increasing function on $\mathbb{R}$. If $u$ is a nonnegative function on $\mathbb{R}$, vanishing at infinity the
    $$
    \int_{\mathbb{R}^{n}} V(|x|)|u^{*}(x)|^{2}dx \leq \int_{\mathbb{R}^{n}}V(|x|)|u(x)|^{2}dx
    $$
\item[(vii)] Fractional Polya-Szeg\"o inequality: For $u\in H^{\alpha}(\mathbb{R})$,   we have 
\begin{equation}\label{PS}
\int_{\mathbb{R}}\int_{\mathbb{R}} \frac{|u^{*}(x + z) - u^{*}(x)|^{2}}{|z|^{1+2\alpha}}dzdx \leq \int_{\mathbb{R}}\int_{\mathbb{R}} \frac{|u(x + z) - u(x)|^{2}}{|z|^{1+2\alpha}}dzdx,
\end{equation}
see \cite{FAEL, YP}.  

\end{itemize}

Finally we recall a result proved by Almgren and Lieb in \cite{FAEL}, which is a crucial ingredient to prove our main theorem in the next section. 
\begin{Thm}\label{PZtm3}
 For each $0< \alpha <1$ and each $n\geq 1$, the map
$
\mathfrak{R} :  H^{\alpha}(\mathbb{R})  \to H^{\alpha}(\mathbb{R})$, defined as 
              $\mathfrak{R}u = u^{*}$,
is continuous and, as a consequence, $\mathfrak{R}: H^{\alpha}(\mathbb{R}) \to H^{\alpha}(\mathbb{R})$ is also continuous.
\end{Thm}

Now, our purpose is to prove the symmetry result for (\ref{Eq00}) by using variational methods jointly with a rearrangement argument. 


\noindent
{\bf Proof of Theorem \ref{TM03}.}
Under ($f_{0}$)-($f_{3}$), ($V_{1}$), ($V_{3}$)  we have proved that $I$ satisfies the mountain pass geometry conditions with mountain pass level 
$$
c = \inf_{\gamma \in \Gamma}\sup_{t\in [0,1]}I(\gamma (t)),
$$
where $\Gamma = \{\gamma \in C([0,1], X^{\alpha})/ \gamma(0) = 0, I(\gamma (1))<0\}$.
By definition of $
c$, for any $n\in \mathbb{N}$, there is $\gamma_{n}\in \Gamma$ such that
\begin{equation}\label{SReq11}
\sup_{t\in[0,1]}I(\gamma_{n}(t)) \leq c + \frac{1}{n^{2}}.
\end{equation}
Now, let $\gamma_{n}^{*}(t) = [\gamma_{n}(t)]^{*}$. By the continuity of rearrangements  in $X^{\alpha}$ we have that $\gamma_{n}^{*} \in \Gamma$. Moreover, by the fractional Polya-Szeg\"o inequality and taking into account that $V$ satisfies ($V_5$), we have
$$
I(\gamma_{n}^{*}(t)) \leq I(\gamma_{n}(t)), \;\;\forall t\in [0,1].
$$
So
\begin{equation}\label{SReq12}
\sup_{t\in [0,1]}I(\gamma_{n}^{*}(t)) \leq c + \frac{1}{n^{2}}.
\end{equation}
As above, by Theorem $4.3$ in \cite{JMMW}, there is a sequence $u_{n} \in X^{\alpha}$ and $\xi_{n}\in [0,1]$ such that
\begin{equation}\label{SReq13}
\| u_{n} - \gamma_{n}^{*}(\xi_{n}) \|_{X^{\alpha}} \leq \frac{1}{n},
\end{equation}
\begin{equation}\label{SReq14}
I(u_{n}) \in (c - \frac{1}{n^{2}}, c + \frac{1}{n^{2}}),
\end{equation}
\begin{equation}\label{SReq15}
\| I'(u_{n}) \|_{(X^{\alpha})'} \leq \frac{1}{n}.
\end{equation}
Following the ideas of the proof of Theorem \ref{TM02}, we can show that
$ u_{n} \to  u,$ $I(u) =  c$, $I'(u)u  =  0$ and finally that
\begin{equation}\label{SReq16}
\lim_{n\to \infty}\|u - \gamma_{n}^{*}(\xi_{n})\|_{X^{\alpha}} = 0,
\end{equation}
concluding the proof. $\Box$



\noindent {\bf Acknowledgements:}
This work was  supported by myself.

\end{document}